\documentclass[prb,twocolumn,amsmath,amssymb,floatfix,superscriptaddress]{revtex4}
\usepackage{graphicx}

\newcommand{\be}{\begin{eqnarray}}
\newcommand{\ee}{\end{eqnarray}}
\newcommand{\bfr}{{\bf r}}
\newcommand{\bfq}{{\bf q}}

\newcommand{\tlomega}{\tilde{\omega}}
\newcommand{\tll}{\tilde{l}}

\newcommand{\tlg}{\tilde{g}}

\newcommand{\bew}{\begin{widetext}}
\newcommand{\enw}{\end{widetext}}

\begin{document}
\draft

\title{Pseudospin ferromagnetism in double--quantum-wire systems}

\author{D.-W. Wang}
\affiliation{Department of Physics,
National Tsing-Hua University, Hsinchu, Taiwan 300, ROC}

\author{E.G. Mishchenko}

\affiliation{Department of Physics, University of Utah, Salt Lake
City, UT 84112, USA}

\author{E. Demler}
\affiliation{Physics Department, Harvard University, Cambridge, MA 02138, USA}

\date{\today}

\begin{abstract}

We propose that a pseudospin ferromagnetic 
(i.e. inter-wire coherent) state can exist in a system of
two parallel wires of finite width in the presence of a perpendicular
magnetic field. This novel quantum many-body state appears when the inter
wire distance decreases below a certain critical value which depends
on the magnetic field.  We determine the phase boundary of the
ferromagnetic phase by analyzing the softening of the spin-mode
velocity using the bosonization approach.  We also discuss signatures
of this state in tunneling and Coulomb drag experiments.
\end{abstract}


\maketitle
Ferromagnetism (FM) in low dimensional
itinerant electronic systems is one of the most interesting
subjects in condensed matter physics. As early as in the '60s Lieb and
Mattis \cite{lieb} (LM) has proved that a ferromagnetic state cannot
exist in one-dimensional (1D) system if the electron-electron
interaction is spin/velocity-independent and symmetric with
respect to the interchange of electron coordinates. Therefore,
possible candidates for 1D FM must involve some
nontrivial modification in the band structure and interaction to
avoid the restrictions of LM's theorem. Most of the examples
proposed in the literature \cite{ferro_review1} rely on some
highly degenerate flat bands (or at least systems
with the divergent density of states) and can be
understood as a generalization of Hund's rule \cite{Hund}. The
only exception appears to be a model of finite range hopping
with a negative tunneling energy \cite{ferro_review2}.

From the experimental point of view, however, physical realization
of the 1D FM in thermodynamical limit is still
absent to the best of our knowledge. In two-dimensions
(2D), some of the most intriguing ferromagnetic systems are the
quantum Hall (QH) bilayers at the total filling factor one.
In these systems the flat band structure is
provided by the magnetic field (Landau levels) and clear experimental
evidence of the 2D pseudospin ferromagnetism (PSFM, with the pseudospin
being the layer index)  has
been observed in the tunneling \cite{spielman01} and drag
experiments \cite{2d_drag_exp} several years after theoretical
proposals \cite{QH_review}.

\begin{figure}
\includegraphics[width=8cm]{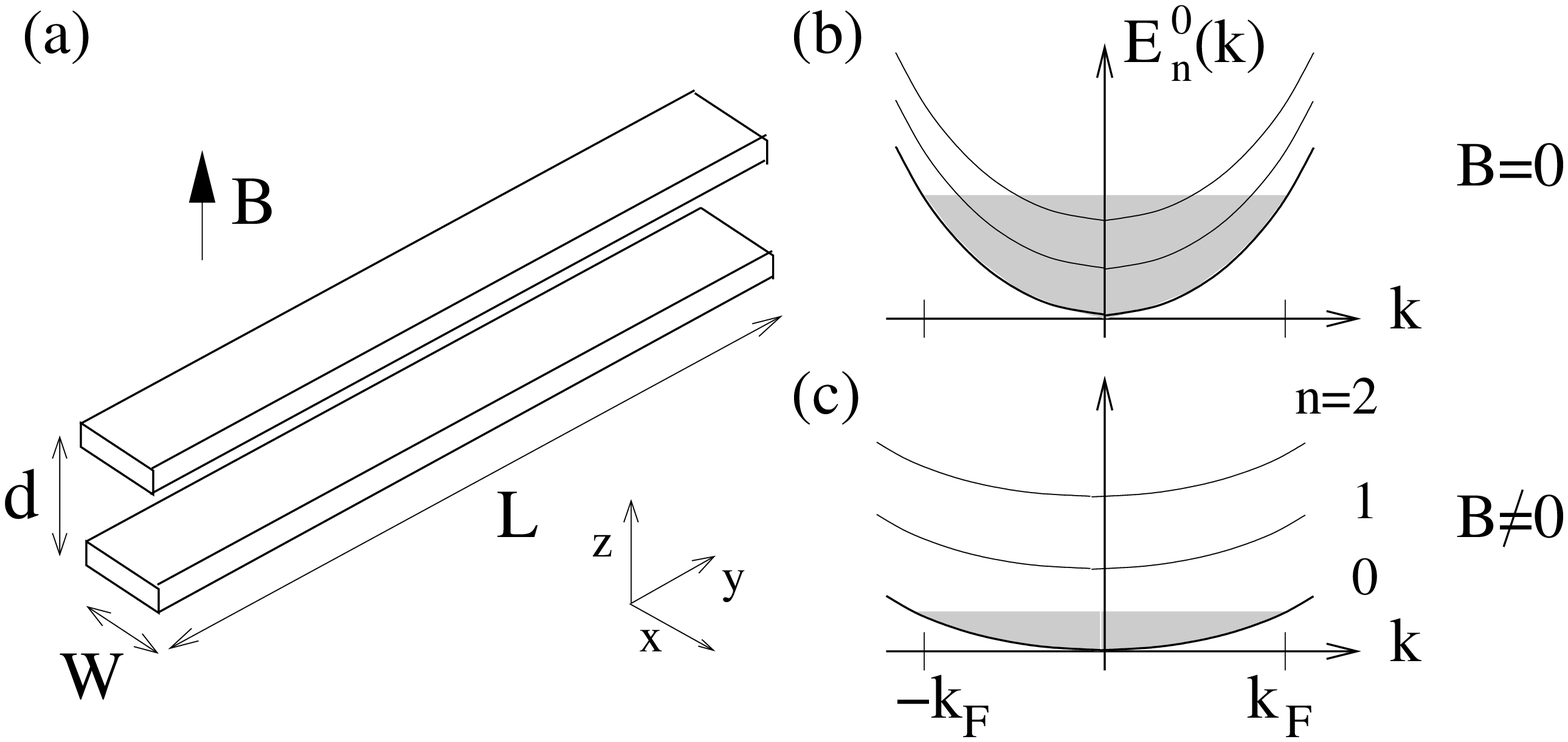}
\caption{ (a) Schematic double wire system considered in this paper.
(b) and (c) are single particle
energy $E^0_n(k)$ of each wire for $B=0$ and $B\neq 0$ cases respectively.
}
\label{system}
\end{figure}

In this paper we propose a realistic {\it one-dimensional} system
which should exhibit a pseudospin 
ferromagnetic order. The system consists of two
{\it finite-width} quantum wires with a magnetic field applied
perpendicular to the wire surface, see Fig.~\ref{system}(a). In the
presence of the magnetic field, single-electron states are modified, which
leads to a strong effective mass enhancement and modification of the
effective Coulomb interaction. These two effects can lead to a 
softening of the spin mode
velocity when the inter-wire distance becomes smaller than a
critical value $d_c$. The system then becomes an easy-plane PSFM state
due to the appearance of interwire coherence (IWC), which
should manifest itself in the appearance
of the resonant peak in the tunneling conductance at small bias
voltages. We also calculate the drag resistance of such
1D PSFM state within the mean-field approximation and
demonstrate that the drag resistance first increases and scales with  
the longitudinal size
as the magnetic field is increased (or the inter-wire distance is 
decreased)  toward
the phase transition boundary and then becomes dramatically reduced
(i.e. not scaled with the size) when entering the PSFM state.
The proposed 1D PSFM transition should be experimentally 
accessible  by the present or near future semiconductor technology.

The double wire system we consider is
aligned in the $y$ direction, Fig.~\ref{system}(a), and centered
at $x=0$ and $z=\pm d/2$. Electrons are confined by a
parabolic potential, $\frac{1}{2}m\omega_0^2x^2$, in the $x$
direction and their motion in $z$ direction is assumed to be
totally quenched. Using the Landau gauge, the single particle
Hamiltonian of momentum $k$ in each wire can be derived to be
\be
H_0&=&-\frac{1}{2m}\partial_x^2+\frac{1}{2}m\tilde{\omega}^2(x-x_0)^2
+\frac{k^2}{2m^\ast},
\label{H_0}
\ee
where $m^\ast=m(\omega_c^2+\omega_0^2)/\omega_0^2$ is the
renormalized electron mass,
$\tilde{\omega}=\sqrt{\omega_0^2+\omega_c^2}$ is the
Landau level splitting, and $x_0=l_0^2k$ is the guiding center
coordinate with $l_0=\sqrt{\omega_c/m(\omega_c^2+\omega_0^2)}$
being the magnetic length. $\omega_c=eB/mc$ is the bare
cyclotron frequency. The wave-functions and energy spectrum of
Eq.~(\ref{H_0}) are easily found from the analogy with the
standard QH system \cite{QH_wang}:
$\psi_{n,k,s}(\bfr)=L^{-1/2}e^{iky}\varphi_n(x+x_0)\sqrt{\delta(z-sd/2)}$
and $E_{n}^0(k)=(n+\frac{1}{2})\tlomega+\frac{k^2}{2 m^\ast}$,
where $n$ is the Landau level index and $s=\pm \frac{1}{2}$ is the
pseudospin index for the upper/lower wire,
$\varphi_n(x)=(\pi^{1/2}2^nn!\, \tll_0)^{-1/2}
e^{-x^2/2\tll_0^2}H_n(x/\tll_0)$ is the $n$-th eigenfunction of a
parabolic potential with $\tll_0\equiv\sqrt{1/m\tlomega}$. 
Throughout this
paper we concentrate on the strong magnetic field (or low electron density)
regime so  that only the lowest energy level ($n=0$) is occupied. One can
see that the magnetic field not only modifies the band
splitting, but also increases the effective mass in the
longitudinal ($y$) direction, leading to a flatband structure with
high density of states similar to the Landau level degeneracy in
2D system, see Fig.~\ref{system}(b).

The interaction Hamiltonian can be derived to be \cite{QH_wang}:
\begin{eqnarray}
H_1&=&\frac{1}{2\Omega_\perp}
\sum_{s_1,s_2,k_1,k_2,{\bfq}_\perp}
V_{s_1,s_2}({\bfq}_\perp,k_1,k_2)
\nonumber\\
&&\times c^{\dagger}_{s_1,k_1+q_y/2}c^{}_{s_1,k_1-q_y/2}
c^{\dagger}_{s_2,k_2-q_y/2}c^{}_{s_2,k_2+q_y/2},
\label{H_1}
\end{eqnarray}
where $c_{s,k}(c^{\dagger}_{s,k})$ are the electron field
operators, $\Omega_\perp=LW$ is the wire area, and 
$V_{s,s'}(\bfq_\perp,k_1,k_2)=A(\bfq_\perp)^2\int\frac{dq_z}{2\pi}
V(\bfq)\left[1+\delta_{s,-s'}(e^{-iq_zd}-1)\right]\,e^{-iq_x(k_1-k_2)l_0^2}$
is the effective 1D interaction with $V(\bfq)$
being the Coulomb interaction. The form-function,
$A({\bfq}_\perp)=\exp\left[-\left(q_x^2\tll_0^{\, 2} 
+q_y^2l_0^4/\tll_0^{\, 2}\right)/4\right]$ is obtained by integrating
the electron spatial wave function \cite{QH_wang}. Due
to the presence of magnetic field, the effective 1D interaction,
$V_{s,s'}(\bfq_\perp,k_1,k_2)$, is  {\it not} equivalent to any
spin-independent (or velocity-independent) symmetric potential.
Thus in our system, the ferromagnetic state is not inhibited by the LM's
theorem.

\begin{figure}
\includegraphics[width=7cm]{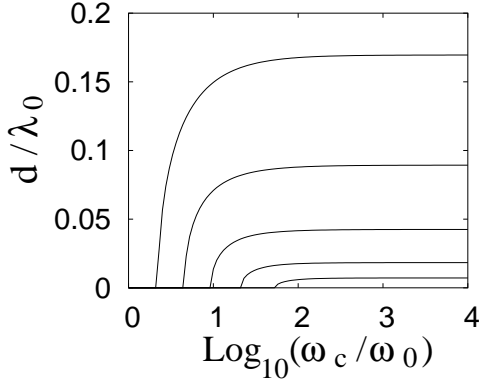}
\caption{ Calculated critical interlayer distance, $d_c$, as a
function of magnetic field ($\propto\omega_c$). 
Electron density in an individual
wire, $n_{e}$, is 0.6, 0.7, $\cdots$, $1.0\times 10^5$ cm$^{-1}$
from top to bottom. Here $\lambda_0=500$ \AA\, and $\omega_0=0.05$ meV.}
\label{dc}
\end{figure}
Starting from Eqs.~(\ref{H_0})-(\ref{H_1}), 
one can use the standard bosonization approach to 
describe the low energy physics near the Fermi points.
After neglecting the irrelevant (nonlocal) terms, we obtain
$H=\sum_{a=\rho,\sigma}H_{a}+H_b$, where
\be
H_{a}&=&\frac{u_a}{2\pi}\int dy\left[
K_a\Pi_a(y)^2+\frac{1}{K_a}\partial_y\Phi_a(y)^2\right].
\ee
Here the sum consists of charge $\rho$ and spin $\sigma$ channels.
$H_b\propto\int dy\cos(\sqrt{8}\Phi_\sigma(y))$ describes the undiagonalizable 
backward scattering term \cite{LL_review}.
$\Pi_a$ and $\Phi_a$ are the bosonic operators satisfying 
the commutation relation: $[\Phi_{a}(y),\Pi_{a'}(y')]=i\delta_{a,a'}
\delta(y-y')$.
The renormalized velocity and Luttinger exponents are
\be
u_a=v_F\sqrt{(1+\tlg_{\theta_a})(1+\tlg_{\phi_a})}, \ \ \
K_a=\sqrt{\frac{1+\tlg_{\theta_a}}{1+\tlg_{\phi_a}}}
\label{uK}
\ee
where $\tlg_{\theta_a/\phi_a}=\frac{1}{2\pi v_F}
\left(2g_{4,a}\mp(2g_{2,a}- g_{1,\|})\right)$ and
$g_{i,\rho/\sigma}\equiv\frac{1}{2}\left(g_{i,\|}\pm
g_{i,\perp}\right)$. Here
$g_{4,\|/\perp}=\int\frac{dq_x}{2\pi}\left[V_{I/O}(q_x,0)\right]$, 
$g_{2,\|/\perp}=\int\frac{dq_x}{2\pi}
V_{I/O}(q_x,0)\cos(2q_xk_Fl_0^2)$, and
$g_{1,\|/\perp}=\int\frac{dq_x}{2\pi}V_{I/O}(q_x,2k_F)$ are defined
as the usual $g$-ology interaction in the Luttinger liquid theory
\cite{LL_review} with $k_F$ being the Fermi momentum. 
$V_I(\bfq_\perp)$ and $V_O(\bfq_\perp)$ are the
intra-wire and inter-wire interaction matrix elements,
respectively. To simplify calculations we model the screened Coulomb
interaction by using $V(\bfq)=(4\pi
e^2\lambda_0^2/\epsilon_0) e^{-|\bfq|^2\lambda_0^2}$ where
$\epsilon_0$ is the static dielectric constant and 
$\lambda_0$ is screening length.
The qualitative results obtained below should not be sensitive to the details 
of the screening potential. 

The ferromagnetic transition occurs as the spin stiffness, $v_{N,\sigma}
=u_\sigma/K_\sigma=v_F(1+\tlg_{\phi_\sigma})$, becomes zero \cite{yang},
or $g_{1\|}=2\pi v_F+2(g_{4,\sigma}+g_{2,\sigma})$.
In general the low energy Luttinger liquid parameters
should be renormalized by the backward scattering, $H_b$,
and therefore the phase
boundary obtained from the bare Luttinger parameters should be modified also.
However, when in PSFM phase, the spin stiffness is negative so that
higher order derivatives, like $\partial_y^2\Phi_\sigma$, has to be included
to stablize the system and to give a nonzero 
spin density, $\rho_s\propto \partial_y\Phi_\sigma$ \cite{yang}.
As a result, the sine-Gordon backward scattering will oscillate
in real space and hence become negligible after averaging in the 
thermodynamical limit. Therefore for simplicity we may assume that 
the renormalization effects are not
very serious so that the phase boundary of the PSFM state can still be
estimated roughly by using the bare Luttinger parameters as stated above.
The critical behavior of similar
transition has been also discussed very recently \cite{yang}.

In  Fig.\ref{dc} we show the calculated critical inter-wire
distance as a function of magnetic field for various single wire electron
densities, $n_e$. PSFM occurs in the
large field and small distance regime. At zero distance,
$g_{2/4,\sigma}=0$, and therefore the critical field
($\omega_{c,cr}$) is the minimum field strength for the
backward interaction ($g_{1,\|}$) to be dominant. On
the other hand, in the extremely large field regime, the Fermi velocity
approaches zero. The critical distance ($d_{cr}$) is now determined
by the competition between the backward scattering and the forward
scattering in the spin channel.
In large density limit ($k_F\lambda_0=\pi n_e\lambda_0>1$) we can
obtain the analytic expression of $\omega_{c,cr}$ and $d_{cr}$:
$\omega_{c,cr}\sim\omega_0\sqrt{2 r_s^{-1}
e^{(2k_F\lambda_0)^2}-1}$ and $d_{cr}\sim \lambda_0
e^{-(2k_F\lambda_0)^2}$, where $r_s\equiv me^2/\epsilon_0\pi k_F$
is the ratio of the average potential and kinetic energies. 
We also checked explicitly that in the 
parameter regime we consider here
the pseudospin polarized state is always energetically unfavorable compared to
the (easy-plane) pseudospin ferromagnetic phase.

We now discuss how such PSFM
phase can be observed in realistic experiments. In this phase 
the system has quasi long-range order
characterized by the presence of a Goldstone mode.
Tunneling spectroscopy has been used to observe similar modes in the
QH bilayers \cite{spielman01} and can be also applied to the present
system. We expect a strong enhancement of the tunneling conductance at
small voltage bias when the system enters the PSFM
state.  Another approach to demonstrating the 1D PSFM in the double
wire system is to perform the Coulomb drag experiments. Such experiments
have been done on 2D\cite{2d_drag_exp} and 1D\cite{1d_drag_exp} 
semiconductor heterostructures in recent
years, and the drag resistance, $R_d$, is a direct measure
of the effects due to inter-wire interaction \cite{Roj}. 
In the literature {\it without} magnetic field or interwire coherence, 
the drag resistance behaves differently in the
two different regimes:
In the perturbative regime $R_d$ vanishes in low temperature limit 
($R_d\propto T^2 \ll e^2/\hbar$) 
\cite{Roj,drag_eugene}; in the strong interaction regime,
however, the backward scattering between the two wires becomes
relevant \cite{NA} and opens a gap $\tilde\Delta$ 
in the energy spectrum, corresponding to the formation of a 
locked charge density wave phase (LCDW) with a divergent drag resistivity $R_d
\propto\exp{(\tilde\Delta/T})$ in low temperature regime.  

To analyze the drag
resistance in the presence of inter-wire coherence, it is useful to employ the
Hartree-Fock (HF) approximation.  This approach neglects long
wavelength fluctuations present in 1D systems, but we expect these
fluctuations give rise only
to small corrections in the drag resistance deep inside the PSFM phase.
The HF Hamiltonian then can be easily diagonalized by transforming the
electron operators into the symmetric ($c^\dagger_{\uparrow,k}
+c^\dagger_{\downarrow,k}$) and the antisymmetric ($c^\dagger_{\uparrow,k}
-c^\dagger_{\downarrow,k}$) channels with the eigenenergies,
$E^\pm_k=k^2/2m^\ast+\Sigma_k\mp\Delta_k-W_0$.
\begin{figure}
\includegraphics[width=6.5cm]{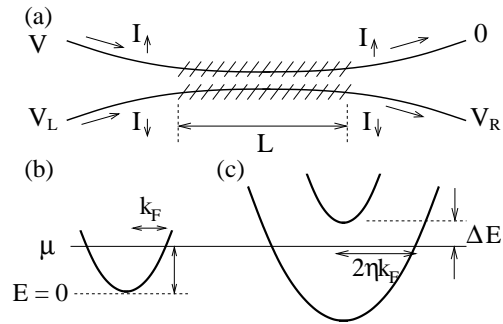}
\caption{
(a) Typical setup for conductance experiment of the
double wire system, where the two wires interact
in the middle regime ($0<y<L$) and are connected to ideal 1D reservoir in
the left ($y<0$) and right ($y>L$) hand sides. The
upper (active) wire is biased by a voltage $V$, while the lower (passive) wire
is biased by $V_R$ and $V_L$ with currents
$I_{\uparrow/\downarrow}$ in the two wires respectively.
(b) and (c) are the band energy for electrons in the incoherent reservoirs and
in the coherent double wire regime respectively. The upper and lower
bands in (b) are for the antisymmetric and symmetric bands respectively.
}
\label{band_shift}
\end{figure}
Here $\Sigma_k$ and $\Delta_k$ are the intra-wire self-energy and the
IWC gap respectively, and $W_0$ is the shift of
the band energy in response to
the reconstruction of the ground state due to coupling to leads,
see Fig. \ref{band_shift}(b)-(c). For simplicity, in our calculation
we neglect the momentum dependenace of $\Sigma_k$ and $\Delta_k$
and approximate them by their values at $k=0$. Within this
approximation, we obtain (at zero temperature):
\be
\Sigma_0 &\sim&
\frac{V_1n_{\rm coh}}{4}(1+e^{-(d/2\lambda_0)^2})
-\frac{V_1}{8\sqrt{\pi}\lambda_y},
\label{Sigma2}
\\
\Delta_0 &\sim& \frac{V_1e^{-(d/2\lambda_0)^2}}{8\sqrt{\pi}\lambda_y}
\label{Delta2}
\ee
where $V_1\equiv e^2\lambda_0/\epsilon_0\lambda_x$,
$\lambda_x\equiv\sqrt{\lambda_0^2+\tll_0^2/4}$, and
$\lambda_y\equiv\sqrt{\lambda_0^2+l_0^4/\tll_0^2}$. $n_{\rm coh}=2n_e$ 
is the total electron density of {\it both} wires in the coherent regime. 
In above equations, we have assumed that
all electrons fall into symmetric band.
This is justified because the bottom of the
antisymmetric band can be shown to be above the chemical potential by
$\Delta E=2\Delta_0-4\eta^2 E_F>0$, when the
magnetic field is large enough (Fermi energy 
$E_F=\frac{k_F^2}{2m^\ast}\propto B^{-2}$).

To calculate the drag resistance in a typical experimental setup,
Fig. \ref{band_shift}(a), we first note that the drag resistance
[$R_d=(V_R-V_L)/I_\uparrow$ for $I_\downarrow=0$] can be expressed
through the conductance of symmetric currents [$G_+=I_\uparrow/V$
for $V_L=V$, $V_R=0$ and hence $I_\uparrow=I_\downarrow$] and the
conductance of antisymmetric currents [$G_-=I_\uparrow/V$ for
$V_L=0$, $V_R=V$ and hence $I_\uparrow=-I_\downarrow$], according
to: $R_d=G_-^{-1}-G_+^{-1}$. The symmetric and antisymmetric
conductances, $G_\pm$, in the presence of inter-wire coherence at
temperature $T$ can be easily derived to be \cite{unpublished},
\be
\left.\begin{array}{c} G_+ \\ G_- \end{array}
\right\}=\frac{e^2}{16\pi T}\int\frac{dE}{\cosh^2\left(\frac{E-E_F}{2T}\right)}
\left\{\begin{array}{l}
|t_s|^2, \\
1-\text{Re}(r_sr_a^\ast), \label{G}
\end{array}
\right.
\ee
where $t_{s/a}$ and $r_{s/a}$ are the transition and reflection
coefficients for the symmetric/antisymmetric channels
respectively. For simplicity we
assume that $\Delta_0$ is  constant for
$0<y<L$ and vanishes outside this interval (the shaded area of
Fig. \ref{band_shift}(a)). We then obtain
\be
\left.\begin{array}{c} t_s \\ r_s \end{array}
\right\}=\frac{1}{D}\left\{\begin{array}{l}
2ik\kappa_s\,e^{-ikL}, \\
(k^2-\kappa_s^2)\sin(\kappa_s L),
\end{array}
\right.
\label{coefficient}
\ee
where $D=(k^2+\kappa_s^2)\sin(\kappa_s L)+2ik\kappa_s\cos(\kappa_s
L)$ and $\kappa_s=\sqrt{k^2+(4\eta^2-1)k_F^2}$. 
The momentum $k$ is related to energy
$E$ in Eq.~(\ref{G}) by $E=k^2/2m^\ast$. $r_a$ is also given by 
Eq. (\ref{coefficient}), replacing $\kappa_s\to i\kappa_a$, where
$\kappa_a=\sqrt{2\xi-(4\eta^2-1)k_F^2-k^2}$ and
$\xi\equiv\Delta_0/E_F$.

\begin{figure}
\includegraphics[width=7cm]{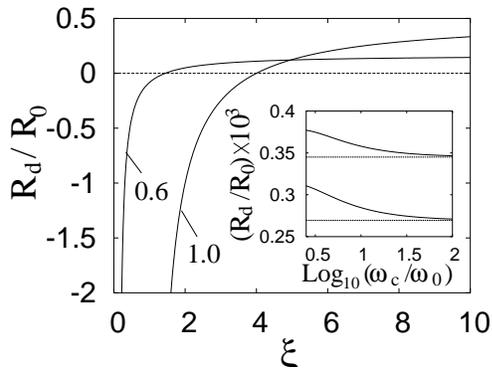}
\caption{Drag conductance as a function of $\xi=\Delta_0/E_F$,
following Eq. (\ref{R_d}).
Results for two electron densities, $\eta$, are
shown together.
Inset: Drag conductance as a function of magnetic field
for $d=0.08\lambda_0$. $n_{\rm res}=0.6$
and $0.7\times 10^5$ cm$^{-1}$ for the lower and upper curves respectively.
}
\label{Rd}
\end{figure}
At zero temperature the conductance and
hence the drag resistance exhibit periodic dependence on the number 
of electrons. At intermediate
temperatures, $v_F/L\ll T\ll E_F$, these oscillations are smeared
out yielding
\be
R_d&=&R_0\left[\frac{(1+2\eta)(2\xi-4\eta^2-1)}
{2\xi+2\eta(1-2\eta)}-\frac{1+4\eta^2}{2\eta}\right],
\label{R_d}
\ee
where $R_0\equiv 2\pi/e^2$. In Fig.~\ref{Rd} we show the
calculated drag resistance as a function of $\xi=\Delta_0/E_F$. It
is negative when $\xi$ is small, but becomes positive with
increasing $\xi$ and eventually saturates at $(1-1/2\eta)R_0$. 

When applying above results to realistic system, one should remember
that due to the repulsive inter-wire interaction, the total electron
density in the coherent regime, $n_{\rm coh}$, should be smaller than
the total electron density in the incoherent wires, $2n_{\rm res}$.
Such electron depletion is negligible in bulk materials due to long
range Coulomb interaction and formation of dipole layers on junction
surfaces. The latter ensure that bringing two bulk 3D materials in
contact and equilibrating their electrochemical potentials does not
change their densities.  In 1D systems, however, the dipole layer
effects are greatly reduced so that the ratio of electron density
inside the IWC regime to the density in the reservoirs, $\eta\equiv
n_{\rm coh}/2n_{\rm res}$, may be appreciable smaller than one.
Within HF approximation, we obtain for small $d$ \cite{unpublished}
\be \eta&=&\frac{1}{2}+
\left(\frac{d}{\lambda_0}\right)^2\frac{(1-1/8\lambda_y
k_F)[1+(\omega_c/\omega_0)^2]}
{16[1+(\omega_c/\omega_0)^2+\lambda_x/4\lambda_0 r_s]},
\label{eta} 
\ee
where $k_F=\pi n_{\rm res}$ is determined by the electron density
in the reservoir.
Using the same parameters as in Fig.~\ref{dc}, we plot the drag
resistance as a function of magnetic field at a given inter-wire
distance and electron density $n_{\rm res}$. 
We note that a finite drag resistance ($R_d$ does 
not scale with the wire length at
$T=0$) is a signature of the coherent state.
The origin of this effect is the indistinguishibility
of electrons flowing in the active and passive wires  ($\langle
c^\dagger_\uparrow c^{}_{\downarrow}\rangle\neq 0$). Similar phenomenon has
already been observed in the 2D QH bilayer systems \cite{2d_drag_exp}.

As mentioned above, without the magnetic field and inter-wire
coherence, the ground state of the
double wire system is predicted to be a LCDW for
long-range Coulomb interaction with an infinite drag resistance
at zero temperature. The effect of
forward scattering could also be relevant\cite{drag_eugene} at
elevated (but still small compared to the Fermi energy)
temperatures. $R_d$ calculated in this scenario always
{\it increases} as the inter-wire
distance decreases, due to the enhancement of inter-wire
interaction. However, as we have shown in this Letter, 
when a strong magnetic field is applied,
a finite $R_d$ that does not scale with the wire length is expected
to be observed when entering the PSFM phase.
Combination of the above two results leads to the following overall
description of the drag reistance:
when the inter-wire distance is decreased from a large value
(or the magnetic field is increased from zero)
the low temperature drag resistance should first increase and reach
a maximum value around the phase boundary (Fig. \ref{dc})
and then begin to decrease to almost zero due to IWC when entering
the PSFM phase. Such
nontrivial behavior of drag resistance could indicate a formation
of 1D pseudospin ferromagnetism in small inter-wire distance or large
magnetic fields.

To summarize, we have shown that in the presence of a strong
magnetic field the electronic system can become (pseudospin)
ferromagnetic in the double quantum wire system. We further
demonstrat that the low temperature 
drag resistance has a non-monotonic behavior
near the phase transition boundary, which should become observable
in the present or near future experiments.

We appreciate fruitful discussion with S. Das Sarma, 
J. Eisenstein, B. Halperin, H.-H. Lin, Y. Oreg, M. Pustilnik, 
A. Shytov, A. Stern, 
A. Yacoby, and M.-F. Yang. This work
was supported by Harvard NSEC and by the NSF Grant DMR02-33773.



\begin{thebibliography}{99}

\bibitem{lieb}
E. Lieb and D. Mattis, Phys. Rev. {\bf 125}, 164 (1962).

\bibitem{ferro_review1}
E. Lieb, Phys. Rev. Lett. {\bf 62}, 1201 (1989);
A. Mielke, J. Phys. A {\bf 24}, L73 (1991);
M. Ulmke, Eur. Phys. J. {\bf B1}, 301 (1998);
T. Okabe, cond-mat/9707032;
L. Bartosch, {\it et al.}, Phys. Rev. B {\bf 67}, 092403 (2003);
H.-H. Lin {\it et al.}, cond-mat/0410654.

\bibitem{Hund}
A. Mielke, Phys. Lett. A {\bf 174}, 443 (1993).

\bibitem{ferro_review2}
H. Tasaki, Phys. Rev. Lett. {\bf 75}, 4678 (1995);
S. Daul and R.M. Noack, Phys. Rev. B {\bf 58}, 2635 (1998), and
reference therein.

\bibitem{spielman01}
I.B. Spielman {\it et al}.,
Phys. Rev. Lett. {\bf 87}, 036803 (2001).

\bibitem{2d_drag_exp}
M. Kellogg, {\it et al.}, Phys. Rev. Lett. {\bf 88}, 126804 (2002);
{\it ibid.}, {\bf 90}, 246801 (2003); {\it ibid.}, {\bf 93}, 036801 (2003);
E. Tutuc, M. Shayegan, D.A. Huse, Phys. Rev. Lett. {\bf 93}, 036802 (2004).

\bibitem{QH_review}
For a review of bilayer QH effect, see
S.M. Girvin and A.H. MacDonald, in {\it Perspectives in Quantum Hall Effects}
edited by S. Das Sarma and A. Pinczuk
(John Wiley \& Sons, New York, 1997); J.P. Eisenstein, {\it ibid}, and
reference therein.

\bibitem{QH_wang}
D.-W. Wang, E. Demler, and S. Das Sarma, Phys. Rev. B {\bf 68}, 165303 (2003).

\bibitem{LL_review}
J. Solyom, Adv. Phys. {\bf 28}, 201 (1979);
J. Voit, Rep. Prog. Phys. {\bf 58}, 977 (1995).



\bibitem{yang}
K. Yang, Phys. Rev. Lett. {\bf 93}, 066401 (2004).

\bibitem{1d_drag_exp}
P. Debray, {\it et al.}, J. Phys. Condens. Matter {\bf 13}, 3389 (2001);
P. Debray, {\it et al.}, Semicond. Sci. Technol. {\bf 17}, R21 (2002);
M. Yamamoto {\it et al.}, Physica {\bf 12E} 726 (2002).


\bibitem{Roj} A. Rojo, J.
Phys.: Condens. Matter {\bf 11}, R31 (1999).

\bibitem{NA} Y.V. Nazarov and D.V. Averin, Phys. Rev. Lett {\bf 81}, 653
(1998); V.V. Ponomarenko and D.V. Averin, Phys. Rev. Lett. {\bf
85}, 4928 (2000); R. Klesse and A. Stern, Phys. Rev. B {\bf 62},
16912 (2000).


\bibitem{unpublished}
E. Mishchenko, D.-W. Wang, and E. Demler, unpublished.

\bibitem{drag_eugene}
M. Pustilnik, {\it et. al.}, Phys. Rev. Lett. {\bf 91}, 126805
(2003).

\end{thebibliography}
\end{document}